\documentclass[runningheads]{llncs}
\usepackage[backend=biber]{biblatex}
\usepackage[T1]{fontenc}
\usepackage[nolist]{acronym}
\usepackage{graphicx}
\usepackage{booktabs}
\usepackage{float}
\usepackage{tabularx}
\usepackage{makecell}
\usepackage[caption=false]{subfig}
\usepackage{siunitx}
\usepackage{listings}
\usepackage[colorlinks=true]{hyperref}
\usepackage{todonotes}
\usepackage{siunitx}

\usepackage{color}

\begin{acronym}[nolist]
    \acro{hpc}[HPC]{High-Performance Computing}
    \acro{simd}[SIMD]{Single-Instruction-Multiple-Data}
    \acro{vlen}[VLEN]{Vector Length}
    \acro{dlen}[DLEN]{Vector Datapath Length}
    \acro{fma}[FMA]{Fused Multiply-Accumulate}
    \acro{rvv}[RVV]{RISC-V Vector Extension}
    \acro{vfpu}[VFPU]{Vector Floating-Point Unit}
    \acro{vlsu}[VLSU]{Vector Load-Store Unit}
\end{acronym}
\begin{document}
\title{Great Expectations: Benchmarking the Real-World Performance of RVV 1.0 in HPC}
%
%
%
\author{Stepan Nassyr\inst{1}\orcidID{0000-0002-0035-244X} \and Prateek Chawla\inst{1}\orcidID{0000-0001-8895-2791} \and Daniel Seibel\inst{1}\orcidID{0000-0003-3833-2089} \and Jayesh Badwaik\inst{1}\orcidID{0000-0002-5252-8179} \and Kaveh Haghighi Mood\inst{1}\orcidID{0000-0002-8578-4961} \and Andreas Herten\inst{1}\orcidID{0000-0002-7150-2505}}
\authorrunning{S. Nassyr et al.}
%
\institute{\textit{J\"ulich Supercomputing Centre} - \textit{Forschungszentrum J\"ulich GmbH}, J\"ulich, Germany}
\maketitle              
\begin{abstract}

Following the ratification of the \ac{rvv} 1.0, new commercially available silicon has been adopting the extension. This paper revisits the question of RISC-V viability for \ac{hpc} by benchmarking the latest \ac{rvv}~1.0-capable hardware (SiFive X280 (Tenstorrent Blackhole), SpacemiT X60 (K1) and X100/A100 (K3), and T-Head C920v2 (Sophon SG2044)). We assess these platforms using standard \ac{hpc} benchmarks (BLAS, FFTW, HPL, HPCG) and synthetic workloads (STREAM, FMA throughput) and compare them to a state-of-the-art HPC ARM64 chip (NVIDIA Grace). Our findings show that while \ac{rvv} 1.0 delivers significant performance improvements over scalar execution, hardware-specific implementation challenges remain. We detail these performance characteristics and discuss the remaining hurdles for RISC-V, including \ac{rvv}, to become a mainstay in the \ac{hpc} landscape.

\end{abstract}
\acresetall
\section{Introduction}


RISC-V is a relatively recent instruction set architecture released under a permissive license that enables royalty-free hardware development.
While it has traditionally been associated with low-power and embedded systems, it has matured significantly over the past decade, with increasing efforts targeting high-performance computing (HPC) systems~\cite{DARE,EUPILOT}.

As part of this evolution, \ac{rvv}~\cite{rvv} was ratified in 2021 and incorporated into the RVA23 profile~\cite{rva23}, providing a standardized foundation for vectorization on future RISC-V systems.
A central design feature of RVV is its vector-length-agnostic (VLA) programming model, which allows applications to be expressed independently of the underlying hardware vector length.
This avoids the portability limitations of fixed-width SIMD extensions such as AVX and AVX-512~\cite{intel_sdm}, where software must explicitly target specific vector widths.
By decoupling software from hardware vector length, RVV enables both portability and scalability across implementations with differing micro-architectural characteristics.

RVV further introduces architectural features that are directly relevant for HPC workloads, including flexible vector register configurations, efficient memory access mechanisms, and support for complex data movement patterns through permutation and gather/scatter operations.
These features are particularly important for computational kernels dominated by dense linear algebra, stencil computations, and memory-bound workloads.

In recent years, the first RISC-V processors targeting HPC workloads with support for the ratified RVV~1.0 specification have started to emerge.
Before ratification, researchers evaluated a range of computational workloads on the proposed RVV 0.7. For instance, Vizcaíno et al. analyzed the performance of multiple applications on a long-vector prototype implementation of RVV~\cite{Vizcaino2023_a,Vizcaino2023_b}. In a more recent study, Banchelli et al. evaluated RISC-V long-vector capabilities for Earth science applications~\cite{Banchelli2026}.
In another study~\cite{Brown2025} evaluated the Sophon SG2044 using the NAS Parallel Benchmark suite, raising the question of whether RISC-V is ready for HPC. 
In this work, we consider four such platforms from different vendors.
The Tenstorrent Blackhole device features a number of SiFive X280 cores, used for supportive tasks for the array of central Tensix cores.
The SpacemiT K1 with its X60 cores is one of the first RISC-V CPUs with RVV support.
Its successor, the SpacemiT K3 featuring X100 and A100 cores is not yet commercially available and was accessed via a cloud-based beta testing environment.
The Sophon SG2044, based on Xuantie C920v2 cores, is the only dedicated, commercially-available server-class RISC-V processor with full RVV~1.0 support at the time of writing and has also been used in prior performance studies~\cite{Brown2025}.

In this work, we revisit the question of RISC-V viability for \ac{hpc} with a focus on numerical libraries enabled for \ac{rvv} and a broader set of RVV~1.0-capable hardware platforms.
We benchmark the considered platforms using a combination of standard \ac{hpc} benchmarks (BLAS, FFTW, HPL, HPCG) and synthetic workloads (STREAM and floating-point throughput tests) enabling a characterization of both computational capabilities and memory subsystem behavior.
Furthermore, we compare these findings to the ARM-based NVIDIA Grace CPU to establish a performance baseline  with a well-known CPU and contextualize the efficiency of the emerging RISC-V platforms within the broader HPC landscape.
The results provide insights into how RVV~1.0 support translates into performance across different implementations and highlight the impact of architectural design choices on observed performance characteristics.

The rest of the paper is structured as follows.
\hyperref[sec:background]{Section~\ref*{sec:background}} provides an overview of the evaluated hardware platforms and the benchmarking methodology.
\hyperref[sec:evaluation]{Section~\ref*{sec:evaluation}} presents the results of both synthetic and standard HPC benchmarks. 
Finally, \autoref{sec:conclusion} summarizes the key findings and outlines directions for future work.


\section{Background}
\label{sec:background}

In this section, we present the devices under test, and the benchmarks used for assessment.

\subsection{Hardware}

\begin{table}[htb]
    \centering
    \caption{Processor Architecture and Performance Comparison. Since the L2 cache is shared for some processor designs, the number of \textit{N} sharing cores is indicated in the form of "/\textit{N}c".}
    \label{tab:processor_specs}
    \scriptsize
    \begin{tabularx}{\linewidth}{X l !{\ } l l l l l l}
        \toprule
        & & \multicolumn{3}{c}{\textbf{SpacemiT}} & \textbf{Sophon} & \textbf{TT} & \textbf{NVIDIA} \\
        \cmidrule(r){3-5} \cmidrule(r){6-6} \cmidrule(r){7-7} \cmidrule(r){8-8}
        \textbf{Metric} & Unit &  \textbf{K1} & \textbf{K3} & \textbf{K3} & \textbf{SG2044} & \textbf{Blackhole} & \textbf{Grace} \\
        \midrule
        \textbf{Arch.} & & X60 & X100 & A100 & C920v2 & X280 & Neoverse V2\\
        \textbf{Release Date} & & 04/2024 & 04/2026 & 04/2026 & 02/2025 & 04/2025 & 05/2023 \\
        \textbf{\#Cores} & & 8 & 8 & 8 & 64 & 4x4 & 72\\
        \textbf{VLEN} & \unit{\bit} & 256 & 256 & 1024 & 128 & 512 & 128\\
        \textbf{DLEN} & \unit{\bit} & 256 & 256 & \makecell[l]{256 (FP64)\\512 (FP32)} & 256 & 256 & 512\\
        \textbf{Peak Perf (FP64)} & \unit{\giga FLOP/\second} & \num{102.4} & \num{140.8} & \num{155.2} & \num{1331.2} & \num{224} & \num{3550} \\
        \textbf{Peak Perf/core (FP64)} & \unit{\giga FLOP/\second} & \num{12.8} & \num{17.6} & \num{14.4} & \num{20.8} & \num{14} & \num{55.3}\\
        \textbf{Peak Perf/core (FP32)} & \unit{\giga FLOP/\second} & \num{25.6} & \num{35.2} & \num{57.6} & \num{41.6} & \num{28} & \num{110.6}\\
        \textbf{L1d Cache} & \unit{\kibi\byte} & \num{32} & \num{64} & \num{64} & \num{64} & \num{32} & \num{64}\\
        \textbf{L2 Cache} & \unit{\kibi\byte} & \num{512}/4c & \num{8192}/8c & \num{8192}/8c & \num{2048}/4c & \num{128}/1c & \num{1024}/1c\\
        \textbf{L3 Cache} & \unit{\mebi\byte} & - & - & - & \num{64}/64c & \num{2}/4c & \num{114}/72c\\
        \textbf{Clock Freq.} & \unit{\giga\hertz} & \num{1.6} & \num{2.2} & \num{1.8} & \num{2.6} & \num{1.75} & \num{3.1} (max. \num{3.456})\\
        \textbf{Frontend} &&  2-way iO & 4-way OoO & 2-way iO & \makecell[l]{3-dispatch/\\4-commit/\\8-issue OoO} & 2-way iO & 6-wide OoO\\
        \bottomrule
    \end{tabularx}
\end{table}

\autoref{tab:processor_specs} shows a selection of relevant architectural details of the processors we evaluate in this work. All of the processors support RVV 1.0, however they differ in the underlying \ac{simd}/vector configurations. The X60, X100 and C920v2~\cite{10.1007/978-3-032-07612-0_44} cores have a \ac{simd}-like setup, with a theoretical throughput of a full \ac{vlen} \ac{rvv} \ac{fma} instruction per-cycle, while the x280 and A100 cores have a different \ac{vlen} and \ac{dlen}, resulting in a full \ac{vlen} instruction to take 2 and 4 cycles respectively. This makes their design closer to a vector unit than \ac{simd}.
The HPC reference system is the NVIDIA's Grace CPU based on the ARM Neoverse V2 architecture supporting ARM's latest SIMD extension, SVE2. Grace was chosen since the SIMD extension SVE is, similar to RVV, vector-length-agnostic and NVIDIA's Grace is currently the best-performing SVE implementation available on the market.

\subsection{Benchmarks}

To evaluate the processors, we use a small set of synthetic and application benchmarks.  These are chosen to cover both peak compute behaviour and more realistic HPC workloads,  as well as to expose potential bottlenecks in the execution and memory subsystems.

\subsubsection{FMA throughput}

We devised a synthetic benchmark that generates assembly loops containing a chain of independent \ac{rvv} \verb|vfmacc| instructions. This allows us to measure the peak computational throughput of each chip for different datatypes. While the application benchmarks we tested use primarily FP64 and FP32 precision, we investigated FP16 throughput for this benchmark, since all of the processors advertise AI focus and it is important to determine whether performance behaviour differs between lower and higher-precision compute.

We added an option to insert \verb|vle| instructions into the \ac{fma} chain at a given ratio. The inserted load instructions use vector registers that aren't used in the \verb|vfmacc| chain and always load from the same address. This allows us to specifically investigate whether there is a bottleneck in the frontend, insufficient L1 port width or some kind of port contention between the \ac{vfpu} and \ac{vlsu}.

\subsubsection{STREAM}
The STREAM benchmark is a synthetic benchmark which measures the sustainable memory bandwidth of a compute node. It only uses little to no computation per byte transferred to or from memory. 

\subsubsection{BLAS}

Basic Linear Algebra Subroutines (BLAS) is a collection of dense linear-algebra routines important for HPC applications. Different implementations of BLAS exist, we focused mainly on the BLIS~\cite{doi:10.1145/2764454} and OpenBLAS~\cite{OpenBLAS} libraries. For BLIS we also created a mechanism that allowed us to override the block sizes $K_C$, $M_C$ and $N_C$ without recompilation and wrote an automatic blocksize optimization tool, with which we determined a highly performant blocksize configuration for each Core.

\subsubsection{FFTW}
FFTW~\cite{FFTW2005} is the most widely used library for discrete Fourier transforms.
Compared to level-3 BLAS and HPL workloads, FFTs are inherently memory bound operations as the arithmetic density is much lower.
For our benchmarks, the RVV~1.0-ready version of FFTW by R. Dolbeau~\cite{dolbeaufftw3} has been used.

\subsubsection{HPL}
The High Performance Linpack (HPL) benchmark~\cite{Linpack} is a widely used dense linear algebra benchmark for evaluating the floating-point performance of HPC systems. It solves a dense system of linear equations using LU factorization with partial pivoting and is the basis for the \textit{TOP500} ranking.

\subsubsection{HPCG}
The High Performance Conjugate Gradient (HPCG) benchmark~\cite{Heroux2013} complements HPL by evaluating performance on workloads representative of sparse iterative solvers. It solves a sparse linear system arising from a three-dimensional discretization using a preconditioned conjugate gradient method with domain decomposition and geometric multigrid preconditioning. In contrast to compute-bound dense linear algebra, HPCG emphasizes memory access patterns, communication, and irregular computation, thereby stressing key system characteristics such as memory bandwidth and network latency.

\section{Evaluation}
\label{sec:evaluation}

In this section, we evaluate the performance characteristics of the considered RISC-V processors using a combination of synthetic and application-level benchmarks. The goal is to systematically analyze both the theoretical capabilities and the practical limitations of the architectures, with particular focus on vector execution, memory hierarchy behavior, and their interaction.

\subsection{Synthetic Benchmarks}
\subsubsection{FMA throughput}\label{sec:fma}
\autoref{tab:processor_fma} shows our results for the FMA throughput benchmark. We have chosen 5 different \ac{fma}/load ratios: only FMAs-baseline, 1, 4, 14, and 26. Ratio 1 represents typical compiler auto-vectorization scenarios. Ratio 4 represents a medium effort manual vectorization for a compute-intensive scenario. Our best-performing GEMM kernel tuned for the SpacemiT X60 has a ratio of 14. Ratio 26 is the maximum possible that can be configured in our tool. 

\begin{table}[hbt]
    \sisetup{table-format = 2.3, table-alignment-mode = format}
    \centering
    \caption{FMA-chain throughput in FLOP/cycle (measured)}
    \label{tab:processor_fma}
    \footnotesize
    \begin{tabular}{l l !{\quad} S[table-format=2.4] !{\,} S !{\,} S[table-format=2.4] !{\ } S[table-format=2.4]!{\ } S[table-format=2.4]!{\ } S[table-format=2.4]}
        \toprule
        & & \multicolumn{3}{c}{\textbf{SpacemiT}} & \textbf{Sophon} & \textbf{TT} & \textbf{NVIDIA} \\
        \cmidrule(r){3-5} \cmidrule(r){6-6} \cmidrule(r){7-7} \cmidrule(r){8-8}
        \textbf{vfmacc/} & \textbf{Data} &  \textbf{K1} & \textbf{K3} & \textbf{K3} & \textbf{SG2044} & \textbf{Blackhole} & \textbf{Grace} \\
        \textbf{vle} & \textbf{Type} & {X60} & {X100} & {A100} & {C920v2} & {x280} & {Neoverse V2} \\
        \midrule
        baseline & \textbf{FP64} & 7.999 & 7.999 & 7.999 & 7.9965 & 7.9871 & 15.9999 \\
        1 & \textbf{FP64} & 2.6624 & 4.517 & 0.5593 & 1.9989 & 3.2171 & 9.9426 \\
        4 & \textbf{FP64} & 5.8561 & 8.0 & 2.6987 & 6.6409 & 7.9873 & 15.9999\\
        14 & \textbf{FP64} & 7.4184 & 8.0 & 7.9238 & 7.9968 & 7.9875 & 15.9999\\
        26 & \textbf{FP64} & 7.4176 & 8.0 & 7.9238 & 7.9973 & 7.9873 & 15.9999\\
        \midrule
        baseline & \textbf{FP32} & 15.997 & 15.999 & 32.985 & 15.9926 & 15.9742 & 31.9998 \\
        1 & \textbf{FP32} & 5.3256 & 9.143 & 1.1441 & 3.9979 & 6.3894 & 19.8889 \\
        4 & \textbf{FP32} & 10.8080 & 15.999 & 5.482 & 13.4759 & 15.9746 & 31.9998 \\
        14 & \textbf{FP32} & 12.3996 & 15.999 & 30.7966 & 15.9918 & 15.9743 & 31.9998 \\
        26 & \textbf{FP32} & 12.3986 & 15.999 & 27.938 & 15.9924 & 15.9742 & 31.9998 \\
        \midrule
        baseline & \textbf{FP16} & 31.9491 & 31.998 & 63.9966 & 31.9778 & 31.9493 & 63.9997 \\
        1 & \textbf{FP16} & 10.6514 & 18.286 & 1.4329 & 7.9962 & 12.779 & 23.9999 \\
        4 & \textbf{FP16} & 21.6148 & 31.998 & 7.9228 & 27.1128 & 31.9479 & 63.9997 \\
        14 & \textbf{FP16} & 24.7973 & 31.998 & 49.4824 & 31.9859 & 31.9482 & 63.9997 \\
        26 & \textbf{FP16} & 24.7967 & 31.998 & 59.4777 & 31.9882 & 31.948 & 63.9997 \\
        \bottomrule
    \end{tabular}
\end{table}

All processors are able to reach their advertised peak performance for each data type when executing a pure \ac{fma} chain. At a 1:1 ratio between \ac{fma} and load instructions, significant slowdown is observed for all processors, the SpacemiT A100 being affected the most, indicating a heavy penalty for frequently switching between load and arithmetic instructions. We observe that the Spacemit X100 and the SiFive X280 cores are able to reach close-to-peak performance with 1 vector load for 4 FMAs. The T-Head C920v2 still exhibits reduced performance at a ratio of 4, but shows close-to-peak performance at 14. The SpacemiT X60 and A100 cores show degraded performance for all ratios, confirming the theory that the \ac{vfpu} and \ac{vlsu} have port contention on these related cores. Out of those two, the A100 struggles much more at ratios of 1 and 4, but looks stronger for high ratios and higher precision reaching up to 99\% for FP64 at a ratio of 14 and 26. This number, while extremely high, is still noticeably behind the X100, C920v2 and X280 cores, indicating that some resource contention remains. NVIDIA Grace shows how mature the Neoverse V2 ARM64 cores are, higher per-cycle performance and reaching it at a ratio of 4.

\subsubsection{STREAM}\label{sec:stream}
The effective memory bandwidth for each chip is measured via \texttt{memcpy} and \texttt{axpy} operations on FP64 data.
Since the conventional STREAM kernels cannot be effectively vectorized for RVV by modern compilers, assembly kernels with explicit RVV~1.0 instructions and a sufficiently high unrolling factor are used to ensure consistent results across the different platforms.
For \texttt{memcpy}, besides the regular implementation with unit-stride vector loads (\texttt{vle64}) and stores (\texttt{vse64}), an implementation using segment vector loads (\texttt{vlseg64}) and stores (\texttt{vsseg64}) is considered.
This is because segment loads and stores are particularly important in applications like FFT with complex floating-point data, where real and imaginary parts can be extracted without additional permutation instructions.
The \texttt{axpy} kernel is implemented in three parts. 
First a read block with unit-stride vector loads (\texttt{vle64}) of \texttt{x} and \texttt{y} data, then a compute block with FMAs (\texttt{vfmacc.vf}) and finally a store section to \texttt{y} with unit-stride vector stores (\texttt{vse64}).
For SVE on Grace, the analogous set of memory instructions (\texttt{ldp}/\texttt{stp} for unit-stride and \texttt{ld2}/\texttt{st2} for segmented) and arithmetic instructions (\texttt{fmla}) is used.

\begin{figure}[!htb]
    \centering
    \subfloat[SpacemiT X60\label{fig:stream-x60}]{%
        \includegraphics[width=0.48\linewidth]{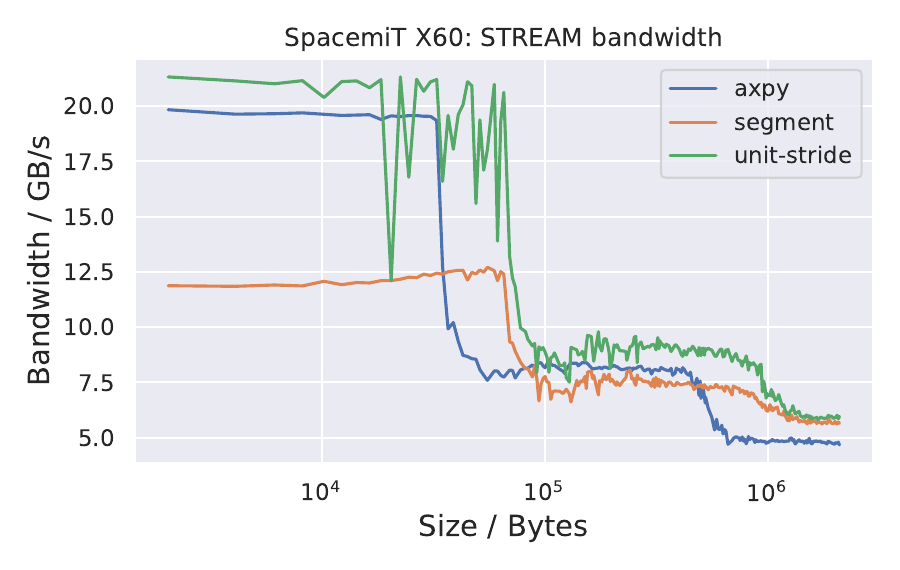}%
    }\hfill
    \subfloat[SiFive X280\label{fig:stream-x280}]{%
        \includegraphics[width=0.48\linewidth]{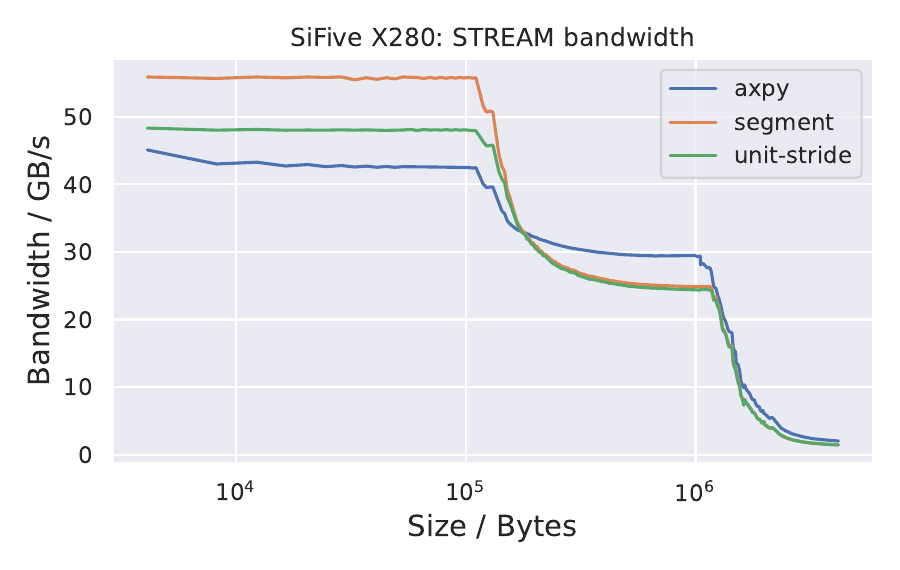}%
    }
    
    \subfloat[SpacemiT X100\label{fig:stream-x100}]{%
        \includegraphics[width=0.48\linewidth]{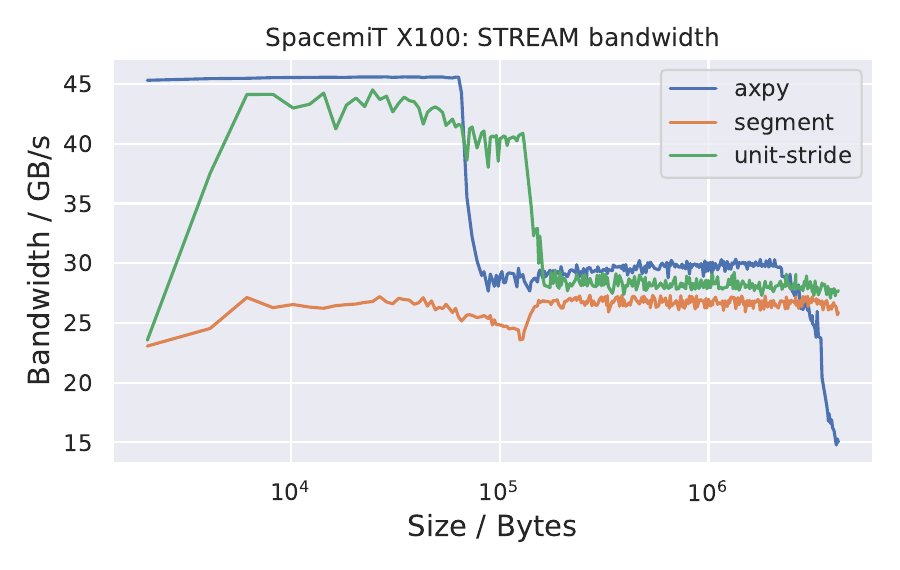}%
    }\hfill
    \subfloat[SpacemiT A100\label{fig:stream-a100}]{%
        \includegraphics[width=0.48\linewidth]{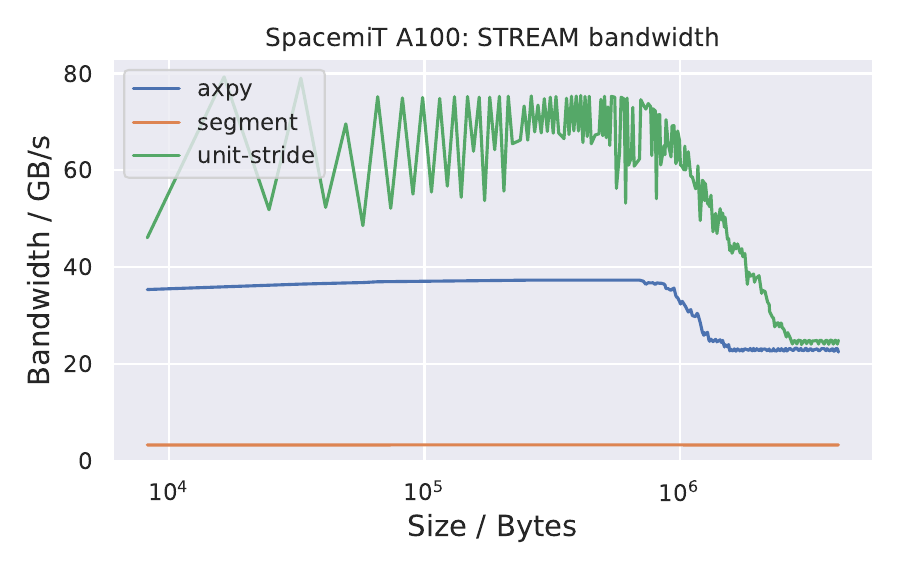}%
    }
    
    \subfloat[Sophon SG2044\label{fig:stream-c920}]{%
        \includegraphics[width=0.48\linewidth]{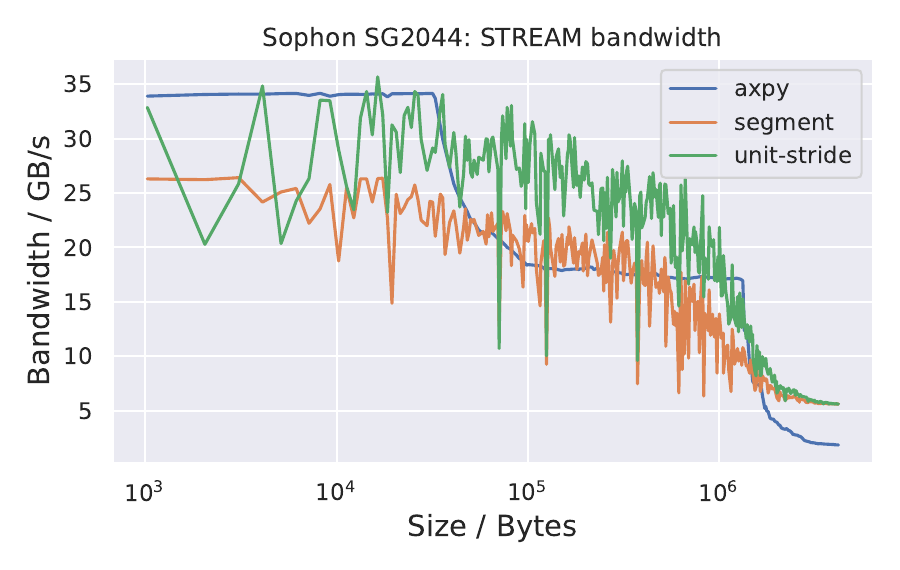}%
    }\hfill
    \subfloat[Grace\label{fig:stream-grace}]{%
        \includegraphics[width=0.48\linewidth]{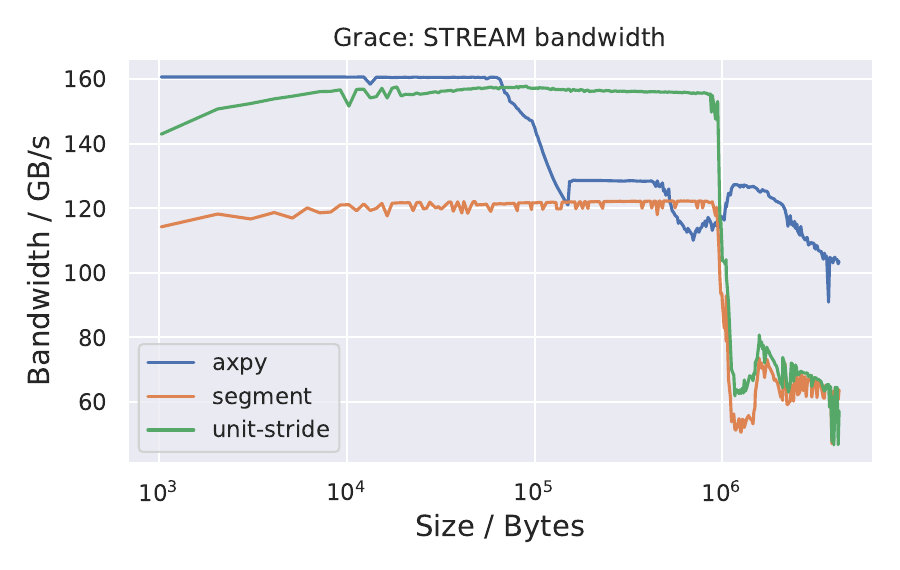}%
    }
    
    \caption{STREAM memory bandwidth benchmarks.}
    \label{fig:combined-stream}
\end{figure}

In Figures~\ref{fig:combined-stream}, the bandwidth (read + write) is reported in GB/s for a single core on each platform.
Note that the $x$-axis is scaled logarithmically, while the $y$-axis is linear.
The peak bandwidth in L1 attainable via regular \texttt{memcpy} is between 22 GB/s for the SpacemiT X60 and 80 GB/s for the SpacemiT A100, which, except for the SiFive X280, is below the theoretical peak bandwidth.
While the SiFive X280 and the SpacemiT X100 are relatively stable in regards to sustained bandwidth, the SpacemiT X60 and A100 as well the Sophon SG2044 experience high fluctuations in L1 and L2 bandwidth depending on the particular message size.
Especially for the Sophon SG2044, the bandwidth can drop by 75\% in certain cases.
Grace reaches a much higher bandwidth with more than 150 GB/s consistently in L1 and L2, which is close to the theoretical maximum.
Given that some of the RISC-V cores like the SpacemiT A100 have a higher theoretical bandwidth per cycle (512 bit vs. 378 bit of Grace), not all of the superior bandwidth of Grace can be attributed to its higher frequency solely.

Except for the SiFive X280, segmented loads and stores are not able to reach the same bandwidth as unit-stride memory instructions.
For the SpacemiT X60 and X100 about 50\% of the bandwidth can be reached, while the Sophon SG2044 maintains about 70\% of the bandwidth.
The SpacemiT A100 experiences a significant slowdown with segment instructions and can only reach a small percentage of bandwidth.
The SiFive X280 on the other hand reaches the theoretical peak of 32 Bytes per cycle in L1 and L2 only with segmented instructions.
For SVE, segmented instructions are called 2-element instructions with de-interleaving and on Grace about 75\% of the peak bandwidth in L1 and L2 can be sustained similar to the Sophon SG2044.

For the \texttt{axpy} kernel, the bandwidth is adequate and remarkably stable for all platforms.
For the SpacemiT X60 and X100 as well the Sophon SG2044, \texttt{axpy} shows only a minor loss or even a small gain of bandwidth in L1 compared to \texttt{memcpy}.
The impact on bandwidth is more pronounced for the SiFive X280 with a 20\% loss in L1 and the SpacemiT A100 with almost 50\% less bandwidth in L1.
In L2, \texttt{axpy} shows similar levels of performance as \texttt{memcpy}.
For Grace the \texttt{axpy} kernel performs exceptionally well, showing that it is able to perform two vector loads, one vector store and one FMA per cycle persistently. Hence even in L3, it is able to reach 100 GB/s of bandwidth.

\subsection{HPC Benchmarks}
\subsubsection{BLAS}
We evaluate BLAS performance on RVV 1.0 CPUs using the BLIS and OpenBLAS libraries and on NVIDIA Grace using NVPL 25.5. For BLIS we evaluate a version previously tuned for the X60 core, as it has proved to perform very well on all processors. For the X280 core specifically, we also tested the upstreamed \verb|sifive_x280| configuration that was contributed by SiFive. For OpenBLAS we chose the Zvl256b configuration for the X100 and x60 cores and Zvl128b for the C920v2. The A100 and X280 were evaluated only with BLIS, as OpenBLAS does not provide a 1024b, 512b or a vector-length agnostic configuration.

In the BLIS framework, block size parameters dictate how matrices are partitioned to optimally utilize the multi-level memory hierarchy (caches and registers) and maximize data reuse. Concurrently, the loop parallelization parameters specify how the nested matrix-multiplication loops are distributed across available threads. When using BLIS with our x60-tuned microkernel we tuned the BLIS block sizes $M_C$,$N_C$ and $K_C$ for each system individually through a rough 3D scan of the parameter space, blocksizes $M_R$ and $N_R$ are hardcoded in the microkernel, which uses 2 vector registers for $M_R$ and 14 elements for $N_R$, i.e. $M_R$ depends on \ac{rvv} VLEN. We also tuned the BLIS loop parallelisation parameters $J_c$,$I_c$,$J_r$ and $I_r$. The threading configuration is shown in \autoref{tab:blis_threading_config}.

\begin{table}[htbp]
    \centering
    \caption{BLIS loop threading parameters for the DGEMM/SGEMM benchmarks.}
    \label{tab:blis_threading_config}
    \begin{tabular}{lcccccc}
        \toprule
        \textbf{Platform} & \textbf{Total Threads} & \textbf{$J_C$} & \textbf{$I_C$} & \textbf{$J_R$} & \textbf{$I_R$} \\
        \midrule
        X60    & 8  & 2 & 1 & 4  & 1 \\
        X100   & 8  & 2 & 1 & 4  & 1 \\
        A100   & 8  & 1 & 1  & 8  & 1 \\
        X280   & 4  &  1 & 1 &  4  & 1  \\
        SG2044 & 64 & 1 & 4 & 16 & 1 \\
        Grace  & 72 & \multicolumn{4}{c}{NVPL (Auto)} \\
        \bottomrule
    \end{tabular}
\end{table}

\begin{figure}[H]
     \centering
         \centering
         \includegraphics[width=\linewidth]{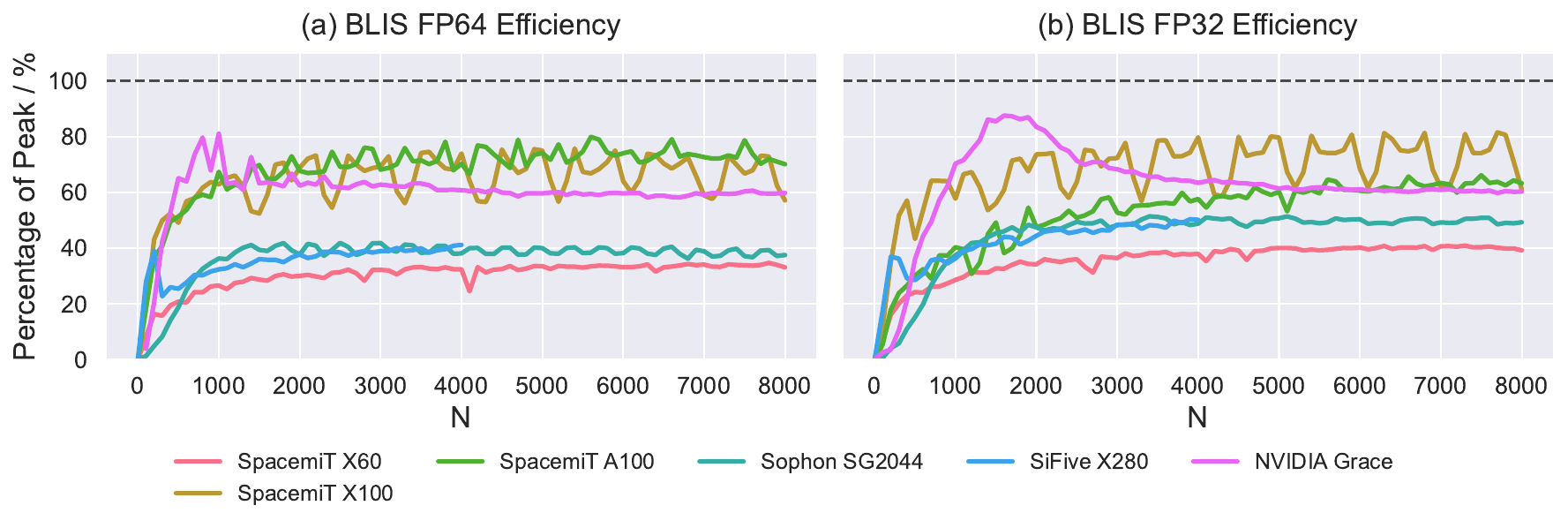}
     \caption{DGEMM and SGEMM efficiency on RVV 1.0 platforms and NVIDIA Grace (ARM64)}
     \label{fig:gemm-efficiency}
\end{figure}

\begin{figure}[!htbp]
    \centering
    
    \subfloat[Double Precision on X60\label{fig:x60_dp}]{%
        \includegraphics[width=0.48\textwidth]{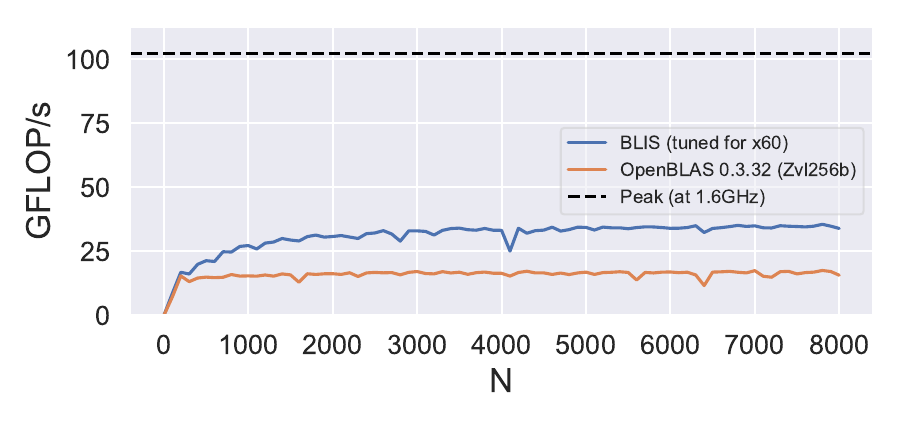}%
    }\hfill
    \subfloat[Single Precision on X60\label{fig:x60_sp}]{%
        \includegraphics[width=0.48\textwidth]{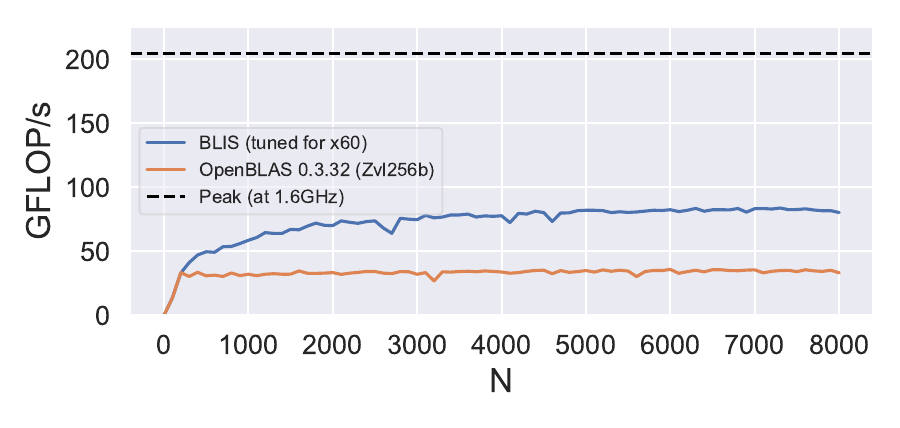}%
    }
    
    \subfloat[Double Precision on X100\label{fig:x100_dp}]{%
        \includegraphics[width=0.48\textwidth]{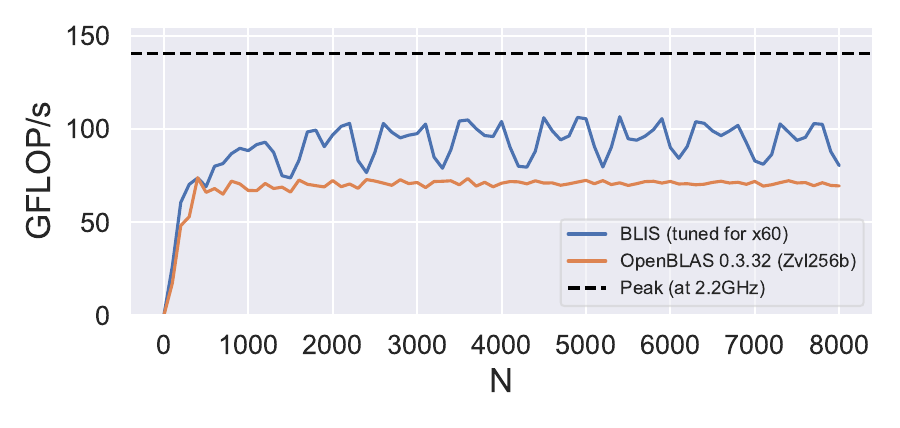}%
    }\hfill
    \subfloat[Single Precision on X100\label{fig:x100_sp}]{%
        \includegraphics[width=0.48\textwidth]{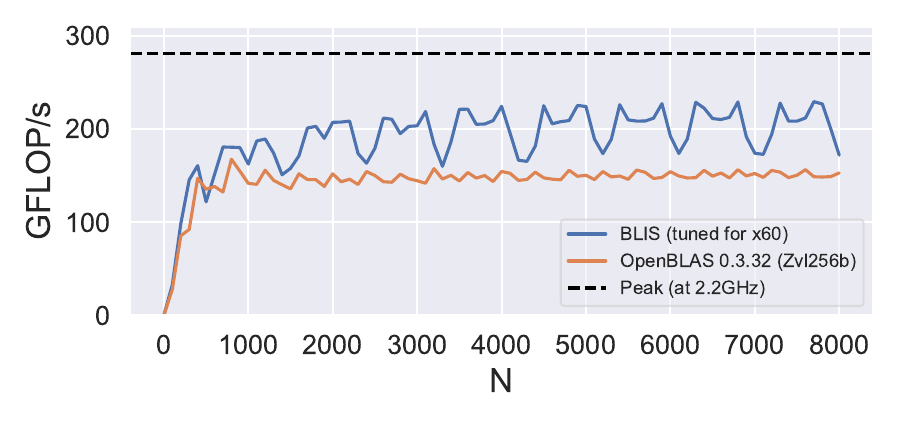}%
    }
    
    \subfloat[Double Precision on A100\label{fig:a100_dp}]{%
        \includegraphics[width=0.48\textwidth]{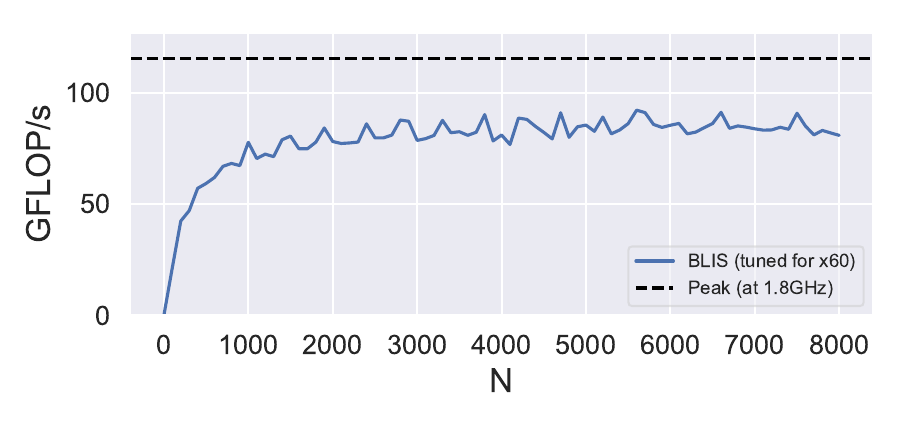}%
    }\hfill
    \subfloat[Single Precision on A100\label{fig:a100_sp}]{%
        \includegraphics[width=0.48\textwidth]{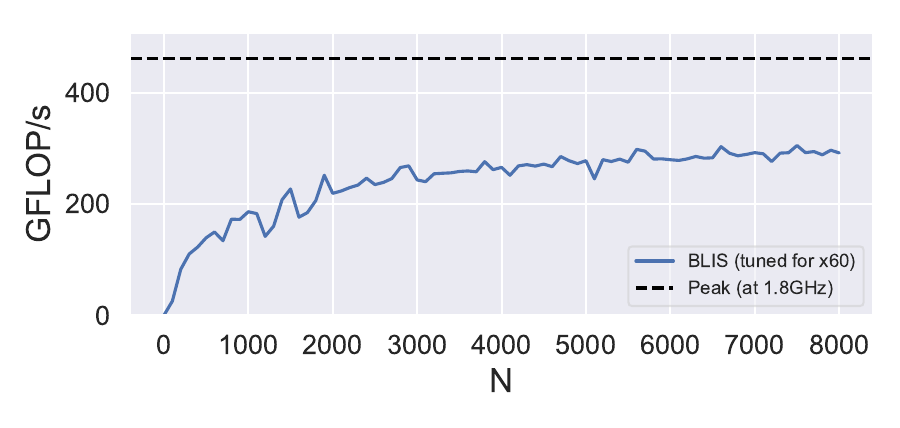}%
    }
    
    \subfloat[Double Precision on SG2044\label{fig:sg2044_dp}]{%
        \includegraphics[width=0.48\textwidth]{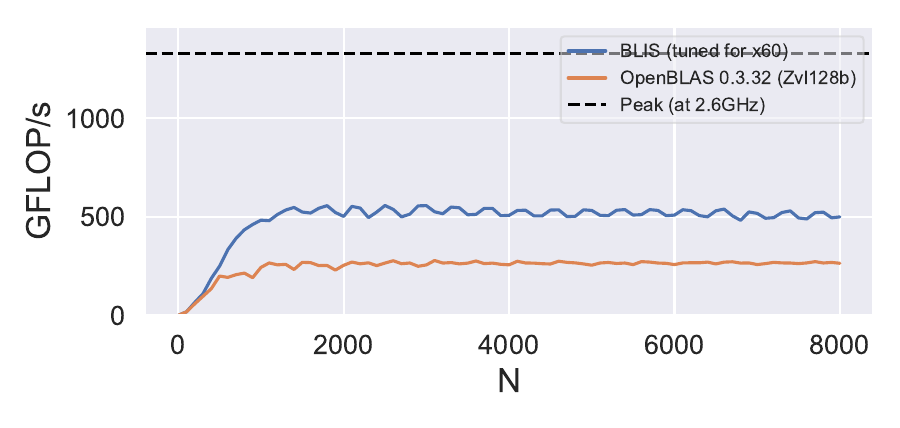}%
    }\hfill
    \subfloat[Single Precision on SG2044\label{fig:sg2044_sp}]{%
        \includegraphics[width=0.48\textwidth]{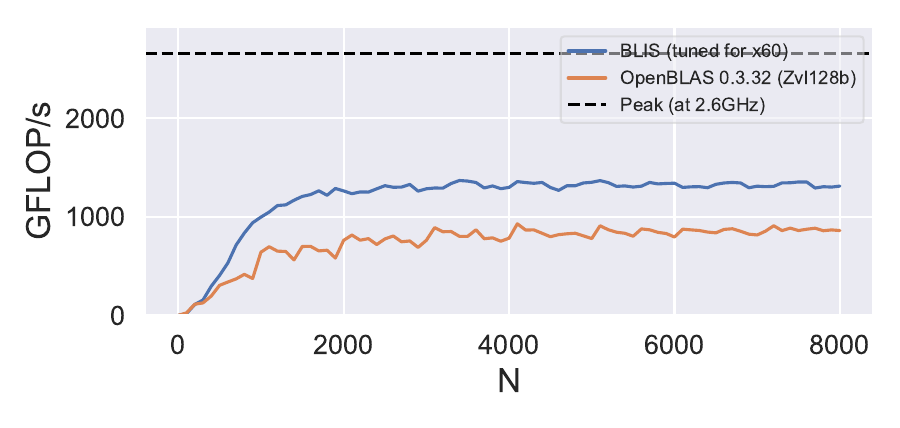}%
    }
    
    \subfloat[Double Precision on X280\label{fig:x280_dp}]{%
        \includegraphics[width=0.48\textwidth]{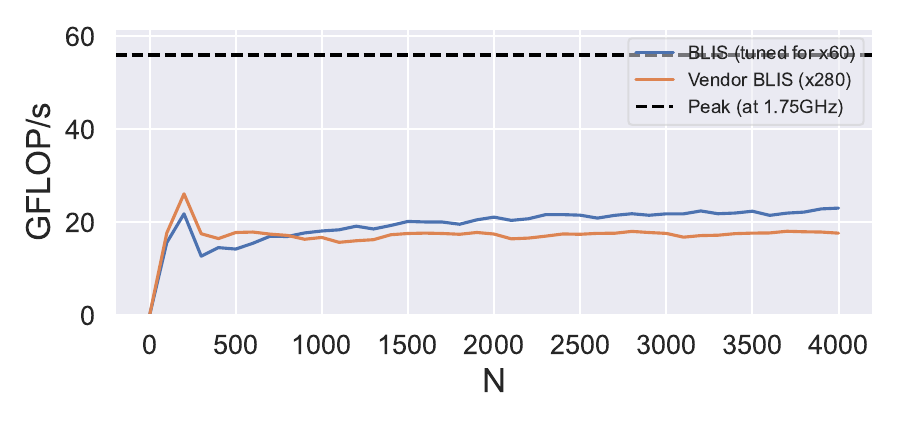}%
    }\hfill
    \subfloat[Single Precision on X280\label{fig:x280_sp}]{%
        \includegraphics[width=0.48\textwidth]{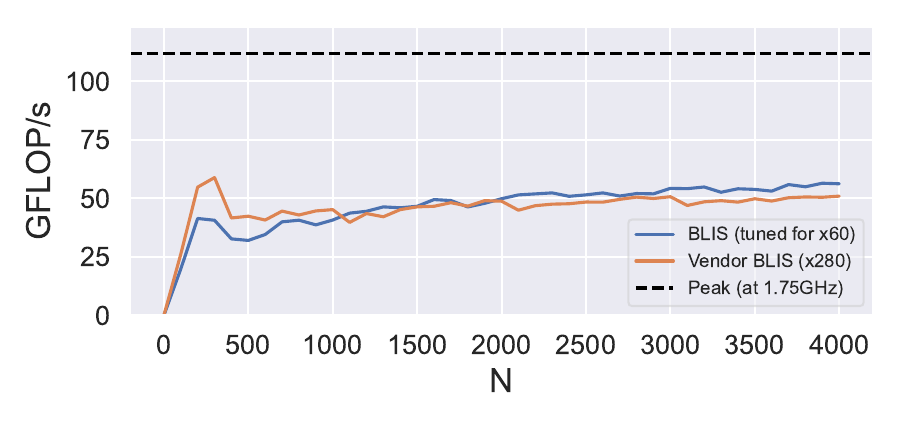}%
    }

    \caption{DGEMM and SGEMM on RVV 1.0 platforms}
    \label{fig:gemm-benchmarks}
\end{figure}

\begin{figure}[!htbp]
    
    \subfloat[Double Precision on NVIDIA Grace\label{fig:grace_dp}]{%
        \includegraphics[width=0.48\textwidth]{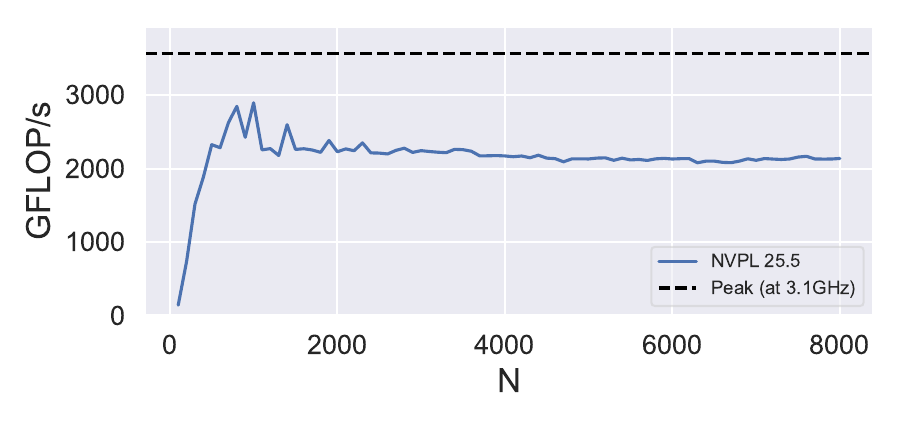}%
    }\hfill
    \subfloat[Single Precision on NVIDIA Grace\label{fig:grace_sp}]{%
        \includegraphics[width=0.48\textwidth]{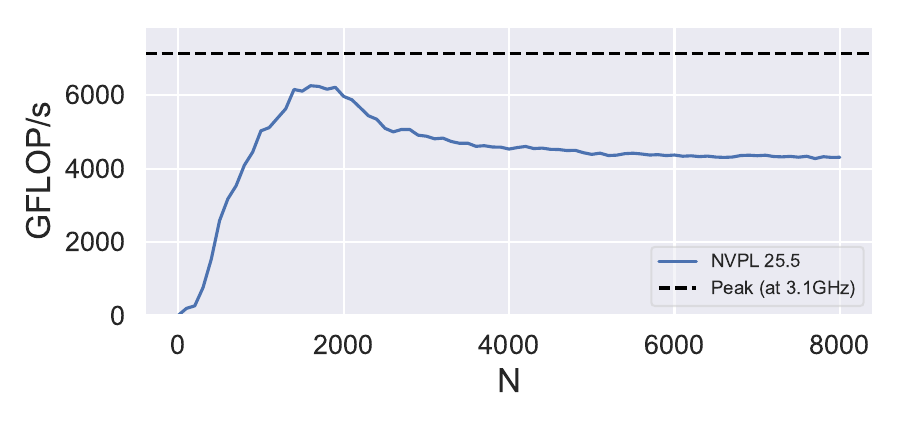}%
    }

    \caption{DGEMM and SGEMM on NVIDIA Grace}
    \label{fig:gemm-benchmarks-grace}
\end{figure}

\hyperref[fig:gemm-benchmarks]{Figure~\ref*{fig:gemm-benchmarks}} show our results for SGEMM and DGEMM on each RVV 1.0 processor, while \autoref{fig:gemm-benchmarks-grace} shows GEMM results on NVIDIA Grace with NVPL 25.5 as an ARM comparison. \autoref{fig:gemm-efficiency} shows the percentage of peak performance reached as a relative efficiency. We observe that while the FMA throughput benchmark shows 80-90\% of peak performance should be attainable with our BLIS GEMM microkernel, the performance is not reached on every processor due to insufficient bandwidth from the memory hierarchy even after a blocksize optimization is performed. This includes the vendor implementation of BLIS for the X280, which in fact is outperformed by our X60-tuned implementation at higher problem sizes. In terms of raw numbers, the 64-core Sophon SG2044 leads, reaching up to \qty{559.6}{\giga FLOP \per \second} for FP64 and \qty{1369.8}{\giga FLOP \per \second} for FP32. The best compute efficiency is achieved with the SpacemiT X100 (75.8\% peak FP64, 81.6\% peak FP32) and A100 (80\% peak FP64, 66.18\% peak FP32) cores. NVIDIA Grace beats all RISC-V platforms in absolute numbers, showing that there is still room for improvement, but in terms of efficiency for large matrices, the A100 and X100 perform competitively, showing microarchitectural strength and potential for a scaled-up version.

\subsubsection{FFTW}
\begin{figure}[H]
    \centering
    \subfloat{%
        \includegraphics[width=0.48\linewidth]{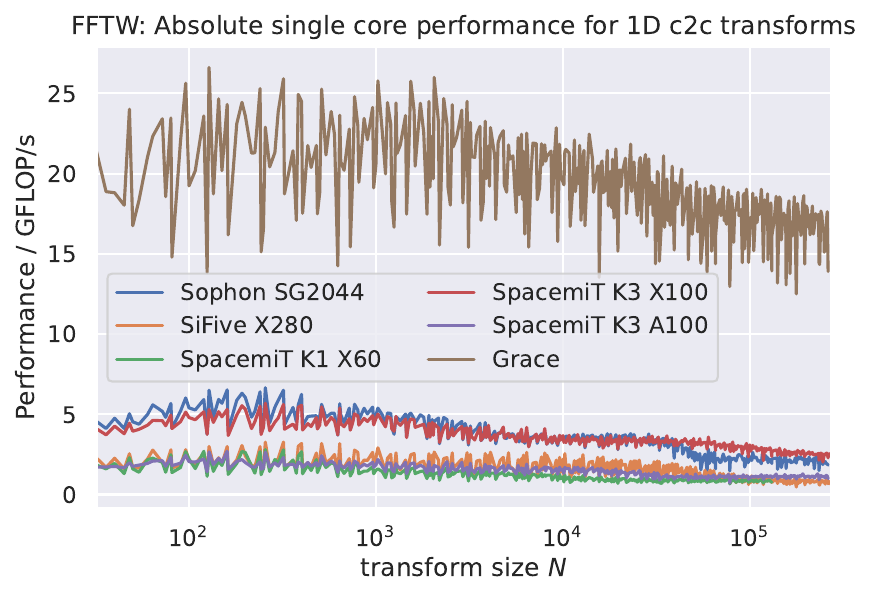}%
    }\hfill
    \subfloat{%
        \includegraphics[width=0.48\linewidth]{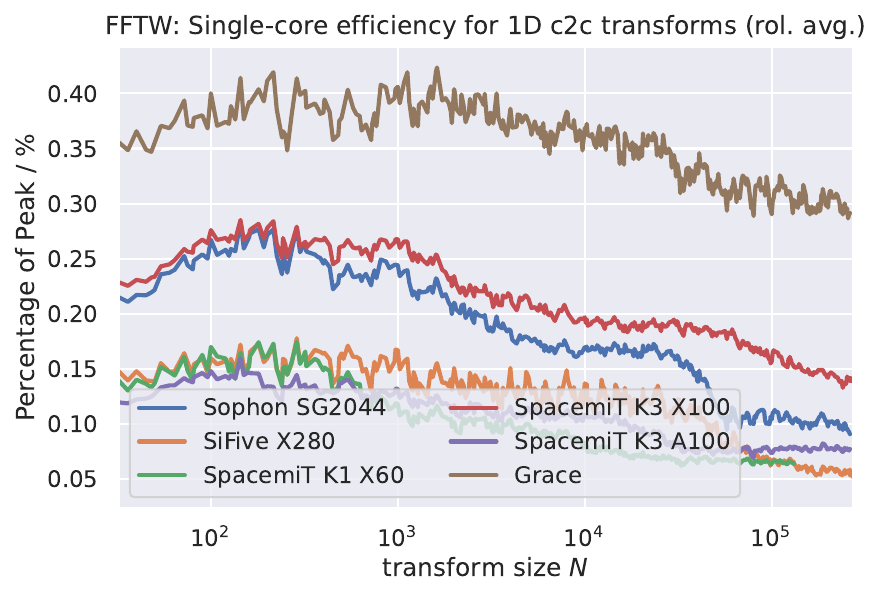}%
    }
    \caption{Single-core performance of FFTW for 1D DFTs}
    \label{fig:fftw}
\end{figure}

The results for FFTW's internal \texttt{libbench2} benchmark in double precision with default planning (\texttt{FFTW\_MEASURE}) are shown in \autoref{fig:fftw} on a single core for each platform.
On the left side, the approximate performance in GFLOP/s for 1D c2c discrete Fourier transforms is plotted against the transform size $N$, while on the right side the performance relative to the peak reported in \autoref{tab:processor_specs} is shown.
For better readability, the right plot shows rolling averages with a window size of $5$.
Note that the GFLOP/s numbers are likely inflated, because FFTW uses the upper bound $5 N \log_2(N)$ instead of the actual number of floating-point operations.
While the Sophon SG2044 leads with about \qty{6}{\giga FLOP \per \second} for sizes $N$ that fit into the L1 cache, the SpacemiT X100 catches up for L2 bound computations, where a performance of \qty{4}{\giga FLOP \per \second} is reached by both. 
Since the SpacemiT X100 has the largest L2 cache, it yields the highest performance for larger $N$ as the other RISC-V systems need to access the L3 cache or RAM.
The other platforms are not able to reach more than \qty{3}{\giga FLOP \per \second}, although the SiFive X280 and SpacemiT A100 have the highest cache bandwidths.
In case of the SiFive X280, this is due to the fact that FFTW's kernels do not implement segment loads and stores despite them being effective for systems like the SiFive X280, see \autoref{sec:stream}.
The SpacemiT A100 exhibits a disproportionate low level performance in accordance with the findings in~\autoref{sec:fma}, where a ratio of at least 14 vector FMAs per load is required to hide latency sufficiently, which is not possible for FFT kernels.
Indeed, a closer look into FFTW's planning output reveals that it falls back to serial non-vectorized codelets in the case of the A100.
Again, Grace is able to outperform all RISC-V platforms by a factor of three to five depending on the transform size and reaches up to \qty{25}{\giga FLOP \per \second}.
However, when measuring in terms of peak performance, the lead is less pronounced with Grace reaching roughly \qty{40}{\percent} compared to a maximum of \qty{28}{\percent} amongst the RISC-V cores.

\subsubsection{HPL}

HPL was evaluated with X60-tuned BLIS kernels for the SpacemiT X60, X100 and A100, the SiFive X280, and the Sophon SG2044 processors.
Although an upstreamed \texttt{sifive\_x280} configuration is available for the X280, our choice was driven by the higher performance observed with the x60-tuned kernels.
For NVIDIA's Grace chip, the specific implementation of HPL shipped by NVIDIA in the HPC benchmarks container v26.02 \cite{nv_hpl} is used. The MPI/BLIS threading configuration for the benchmarks is shown in \autoref{tab:threading_config}.

\begin{table}[htbp]
    \centering
    \caption{Multi-threaded configuration and BLIS loop threading parameters for the HPL benchmark}
    \label{tab:threading_config}
    \begin{tabular}{lcccccc}
        \toprule
        \textbf{Platform} & \textbf{Total Threads} & \textbf{MPI Ranks} & \textbf{$J_C$} & \textbf{$I_C$} & \textbf{$J_R$} & \textbf{$I_R$} \\
        \midrule
        X60    & 8  & 2 & 1 & 1 & 4  & 1 \\
        X100   & 8  & 2 & 1 & 1 & 4  & 1 \\
        A100   & 8   & 1  & 1  & 1  & 8   &  1 \\
        X280   &  4  &  1 &  1 & 1  & 4   & 1  \\
        SG2044 & 64 & 4 & 1 & 1 & 16 & 1 \\
        Grace  &  72 & 4  & \multicolumn{4}{c}{NVPL (Auto)} \\
        \bottomrule
    \end{tabular}
\end{table}

\begin{figure}[hbt]
    \centering
        \includegraphics[width=0.90\linewidth]{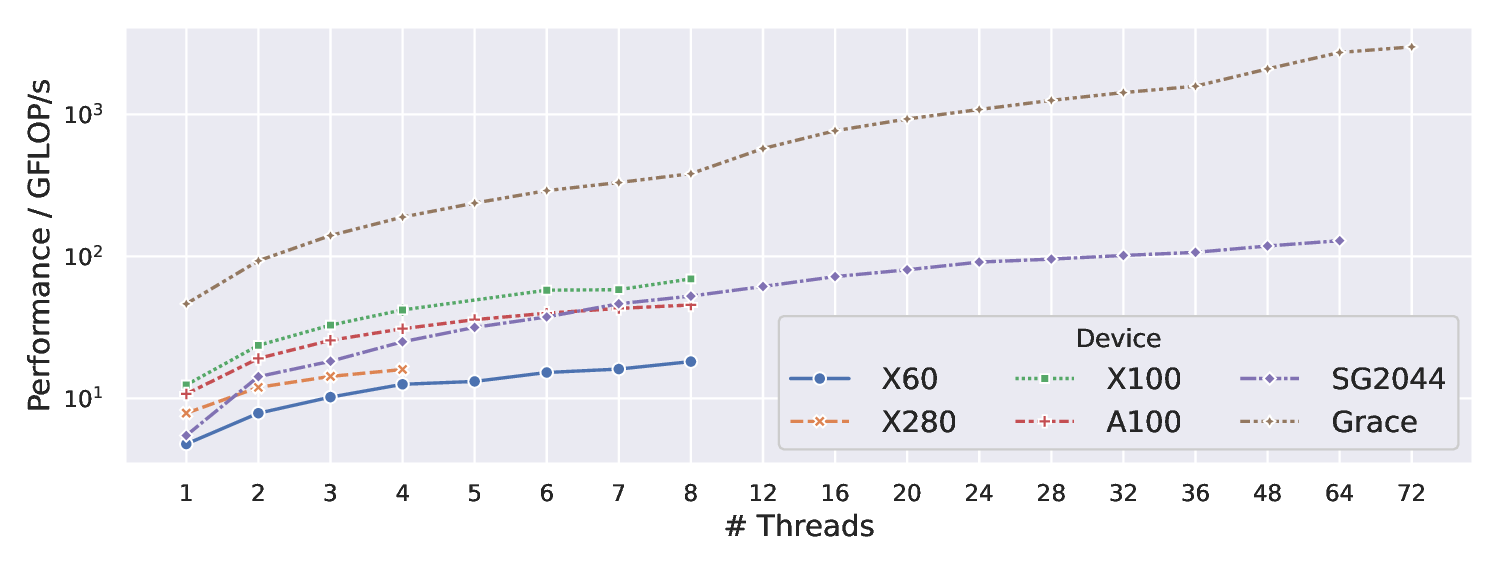}%
    \caption{Performance characteristics in the HPL benchmark.}
    \label{fig:hpl-results}
\end{figure}

\autoref{fig:hpl-results} shows the results of our benchmarks on each processor. The Sophon SG2044 displays the least single-threaded performance at \qty{5.48}{\giga FLOP \per \second}, with \qty{6.48}{\giga FLOP \per \second} for the raw DGEMM calculation, and about \qty{12.2}{\percent} of the total time is spent in panel factorization. 
The SpacemiT X100 cores show the best multi-threaded performance overall, with the 8-threaded HPL achieving 83.64 GFLOP/s in raw DGEMM, while spending roughly \qty{12.15}{\percent} of the total time in panel factorization. This is in contrast to the SG2044 chip, which at maximum threading (in this case, 64 threads) achieves \qty{417.43}{\giga FLOP \per \second} at raw DGEMM, but the overall performance is severely impacted due to it spending \qty{54.22}{\percent} of the time in panel factorization. The RISC-V chips tested are completely outmatched by the NVIDIA Grace CPU which achieves \qty{46.37}{\giga FLOP \per \second} in single-threaded HPL, and with 24 total threads (across 2 MPI ranks) about \qty{1084} {\giga FLOP \per \second} are reached in testing.


\subsubsection{HPCG}

We evaluated the SpacemiT X60, X100, A100 and the SiFive X280 with the reference kernels in the HPCG benchmark. The reference kernels simulate a ``good-enough'' parallelized load expected from a (portable) application, providing a good starting point for a real-world scientific workload. In order to compare to a commercially available in-production chip, we also present results for the NVIDIA Grace CPU. The HPCG benchmark was run via the available NVIDIA container \cite{nv_hpl}.

\begin{figure}[hbt]
    \centering
    \subfloat[Raw FP performance\label{fig:hpcg-perf}]{%
        \includegraphics[width=0.48\linewidth]{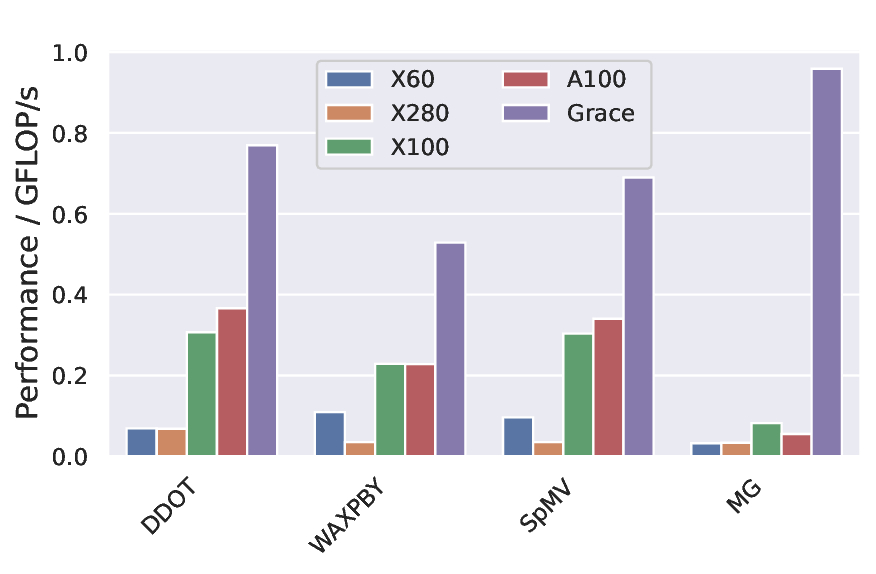}%
    }\hfill
    \subfloat[Raw Memory bandwidth\label{fig:hpcg-bw}]{%
        \includegraphics[width=0.48\linewidth]{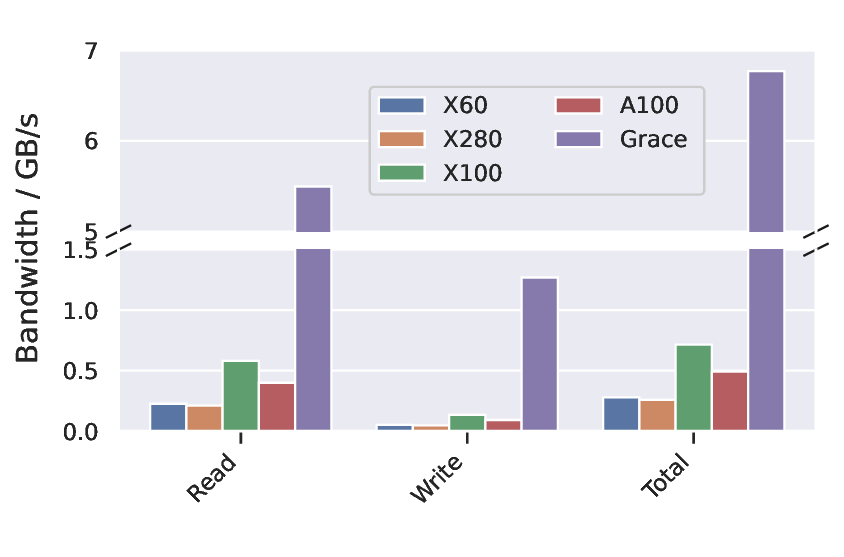}%
    }
    \caption{Compute and memory performance (per-core) on reference HPCG kernels.}
    \label{fig:hpcg-results}
\end{figure}

\autoref{fig:hpcg-results} shows the results of our benchmarks on each processor. Memory bandwidth limitations affect the X280 processor quite severely, which is seen in memory-intensive kernels like the \lstinline{WAXPBY} ($\vec{w} \gets \alpha \vec{x} + \beta \vec{y}$) and the \lstinline{SpMV} (sparse matrix-vector multiplication). The A100 cores show a higher compute performance than the X100 cores, which is a complete reversal of the HPL results from \autoref{fig:hpl-results}. This is not the case for \lstinline{WAXPBY}, where the higher clock speed of the X100 still allows it to take the top spot. Overall, the SpacemiT A100 cores provide the best performance for reference HPCG kernels. We also see significant inter-generational improvement between the X100 and its predecessor, the X60. The X100 improves its DDOT performance by a factor of \qty{3.43}{}, and the SpMV by a factor of \qty{2.15}{}. The total memory bandwidth is also improved by a factor of \qty{1.56}{}.

However, a real-world comparison to an established microarchitecture (ARM, in this case) exposes a significant performance gap. The A100, the top performer for DDOT, achieved \qty{0.34}{\giga FLOP \per \second} per core, which is roughly half the performance of the Grace chip (\qty{0.77} {\giga FLOP \per \second}). This gap is much higher in case of memory bandwidth, where the highest performing chip among the RISC-V lineup, the X100, achieves \qty{0.49} {\giga \byte / \second} per core compared to the more optimized Grace CPU, which achieved \qty{6.77} {\giga \byte / \second} per core in our testing runs. Data from the Sophon SG2044 could not be obtained due to the cluster being down for maintenance.


\section{Conclusion}
\label{sec:conclusion}
In this work, we revisited the question of RISC-V's readiness for \ac{hpc} by evaluating five \ac{rvv} 1.0-capable processors, with a specific focus on the performance of \ac{rvv}-accelerated numerical libraries. Our synthetic compute and memory throughput benchmarks exposed the physical limits of current hardware, revealing micro-architectural bottlenecks such as execution port contention and memory-hierarchy constraints that frequently prevent workloads from fully utilizing the compute pipelines.

These architectural limitations become apparent in our BLAS evaluation, where most platforms struggle to overcome the memory wall, yielding compute efficiencies between 30\% and 50\%. The SpacemiT K3, however, emerges as a major exception and represents a profound generational leap over the preceding K1. By significantly improving its vector front-end and memory bandwidth, the K3 achieves compute efficiencies of up to 80\% on both the X100 and A100 cores. This translates directly into macro-benchmark dominance, with the K3 outperforming the competition in sustained HPL and HPCG throughput. Conversely, our FFTW evaluation highlights the distinct advantages of out-of-order execution, favoring the Sophon SG2044 and SpacemiT X100 cores for complex memory access patterns.

Furthermore, broader cross-architecture comparisons with established devices like NVIDIA's ARM-based Grace CPU reveal that a substantial performance gap still remains, which go beyond differences in clock frequency and advantages in manufacturing fabrication.
However, the rapid inter-generational improvements observed across the evaluated RISC-V designs strongly indicate that this performance gap is poised to close.

Ultimately, we observe a clear and rapidly accelerating path to the viability of \ac{rvv} hardware in \ac{hpc}. Resolving the remaining readiness gaps requires a dual approach: aggressively scaling out core counts, a paradigm proven highly capable by the 64-core Sophon SG2044, while simultaneously eliminating internal pipeline and bandwidth bottlenecks, as demonstrated by the SpacemiT K3. As these design philosophies mature and converge, RISC-V with its RVV extension is positioned to become a genuinely competitive force in the \ac{hpc} landscape.
\begin{credits}
\subsubsection{\ackname} 
Funding for parts of this work has been received from EuroHPC's project DARE SGA 1 under Grant Agreement No.~101202459. The authors gratefully acknowledge the Monte Cimone project for providing access to the Sophon compute nodes used in this work. 
\subsubsection{\discintname}


\end{credits}

\clearpage

\printbibliography

\end{document}